\documentclass[12pt]{article}
\usepackage{amssymb,amsmath,epsfig,cite}
\allowdisplaybreaks

\begin{document}
\title{\bf Gravastars in $f(R,G)$ Gravity}

\author{ M. Z. Bhatti \thanks{mzaeem.math@pu.edu.pk},
Z. Yousaf \thanks{zeeshan.math@pu.edu.pk}, and A. Rehman \thanks{attiqwatto786@gmail.com}\\
Department of Mathematics, University of the Punjab,\\
Quaid-i-Azam Campus, Lahore-54590, Pakistan.}

\date{}

\maketitle
\begin{abstract}
In this paper, we discuss some feasible features of gravastar that
were firstly demonstrated by Mazur and Mottola. It is already
established that gravastar associates the de-Sitter spacetime in its
inner sector with the Schwarzschild geometry at its exterior through
the thin shell possessing the ultra-relativistic matter. We have
explored the singularity free spherical model with a particular
equation of state under the influence of $f(R,G)$ gravity, where $R$
is the Ricci scalar and $G$ is the Gauss-Bonnet term. The interior
geometry is matched with a suitable exterior using Israel formalism.
Also, we discussed a feasible solution of gravastar which describes
the other physically sustainable factors under the influence of
$f(R,G)$ gravity. Different realistic characteristics of the
gravastar model are discussed, in particular, shell's length,
entropy, and energy. A significant role of this particular gravity
is examined for the sustainability of gravastar model.
\end{abstract}

\section{Introduction}

A gravastar, abbreviated for gravitationally vacuum star, is a
celestial object contemplated as an alternative to black hole
structure. Mazur and Mottola \cite{mazur2004gravitational} extended
the Bose-Einstein condensate concept to the gravity system and
devised a cold compact object that is named as gravastar whose
radius is equivalent to the Schwarzschild radius. Gravastar does not
contain an event horizon or an essential singularity and is
considered as feasible alternate to black hole. A shell of
ultra-relativistic fluid rings the inside of gravastar while the
outside sector is completely a vacuum. The shell is very little in
thickness having the width within the range of $r_{2}> r > r_{1}$,
here $r_{1}\equiv D$ while $r_{2}\equiv D+\epsilon$ are the radii
representing the inner and outer sectors of gravastar. The equation
of state (EoS) used for the description of complete structure of
gravastar has
\begin{enumerate}
\item Inner sector $(r_{1}>r\geq0 ):~ p=-\rho$,
\item Shell $(r_{2}\geq r\geq r_{1}):~p=\rho$,
\item Outer sector $(r>r_{2}):~p=\rho=0$.
\end{enumerate}
The inner sector of the gravastar acts like dark energy applying the
opposite effects to gravitating force in the thin shell and avoids
the formation of singularity. The outer region is totally vacuum and
could be elaborated by the Schwarzschild model.

A plenty of work associated with the mathematical and the physical
problems of gravastar is accessible in literature. These works are
mostly accomplished in Einstein's general relativity
\cite{lobo2013linearized,nandi2009energetics,horvat2007gravastar,debenedictis2006gravastar,bilic2006born,visser2004stable,lobo2007gravastars,rocha2008stable,
bhatti2019locally,bhatti2020charged}. However, their results are
then modified to various modified theories
\cite{yousaf2019charged,yousaf2020gravastars,yousaf2020construction}.
The presence of dark matter and the concept of escalating universe
has challenged the theory of general relativity
\cite{riess1998observational,perlmutter1999measurements,de2000flat,peebles2003cosmological}.
To elaborate the cosmological events in the reign of the strong
field, few alterations are needed in general relativity (GR),
consequently, various theories are recommended. The same idea came
in the mind of Buchdahl and he proposed $f(R)$ theory
\cite{buchdahl1970non} in 1970. A consolidated cosmic history in
$f(R)$ theory is explained by Nojiri and Odintsov
\cite{nojiri2011unified}. Baibosunov \cite{baibosunov1990model} has
elaborated the early universe model in $f(R)$ theory. Harko
\cite{harko2011evolution} conferred the $f(R,T)$ gravity by
including the contribution due to matter source. Also, the
Guass-Bonnet theory \cite{lovelock1971einstein} has further provided
the alteration to GR adding the term of Guass-Bonnet $(G \equiv
R^{2}-4R_{\alpha\beta}R^{\alpha\beta}+
R_{\alpha\beta\rho\sigma}R^{\alpha\beta\rho\sigma})$ in the
Einstein-Hilbert action where $ R_{\alpha\beta}$ represent Ricci
curvature tensor and $R_{\alpha\beta\rho\sigma}$ indicates the
Riemann curvature tensor..

Bamba \emph{et al.} \cite{bamba2017energy} suggested the $f(G)$
theory and discussed the energy conditions for renowned FLRW metric.
Some thought provoking work can be seen in this alternative theory
\cite{shamir2017noether,shamir2017some}. It is contended that the
$f(R,G)$ gravity can be encountered as a suitable alternative
gravity for GR as the effective formation. We can note the
application of $f(R,G)$ theory to distinct cosmological fields
\cite{li2007cosmology,lattimer2014neutron,jaffe2011connecting,de2009construction,alimohammadi2009remarks,boehmer2009stability,uddin2009cosmological}.
The stability of relativistic interiors as well as the existence of
self-gravitating structures has been discussed widely in literature
\cite{bhatti2020electromagnetic,yousaf2020new,bhatti2020stability}.

Felice and Tanaka \cite{de2010inevitable} studied the the dynamical
features of anisotropic cosmic evolution with the help of linear
perturbation theory by taking some generic forms of $f(R,G)$ models.
Felice \emph{et al.} \cite{de2011stability} analyzed the stability
of Schwarzschild-like solutions in models of $f(R,G)$ theory.
Dombriz and Gomez \cite{de2012stability} have discussed the
stability of the cosmological solutions and elaborated the possible
results of modified gravity on cosmological level. De Laurentis and
Revelles \cite{de2014newtonian} discussed the Newtonian, ppN and pN
restrictions in $f(R,G)$ theory. Houndjo \emph{et al.}
\cite{houndjo2014exploring} presented thorough description of
cylindrically symmetric solutions for $f(G)$ gravity. Atazadeh and
Darabi \cite{Atazadeh2014} studied the viability of $f(R,G)$ gravity
models with the help of energy conditions. De Laurentis \emph{et
al.} \cite{de2015cosmological} has described the cosmological
inflation under the effects of $f(R,G)$ theory. Odintsov \emph{et
al.} \cite{odintsov2019dynamics} explained distinct dark energy and
inflationary enlargements in the scenario of $f(R,G)$ theory.
Recently, Shekh \emph{et al.} \cite{doi:10.1142/S0219887820500486}
calculated few cosmic models in $f(R,G)$ gravity and performed
dynamical analysis in order to study the viability of their results
through energy conditions.

Different astrophysical results including  the renowned $\Lambda
CDM$ model have been elaborated by Myrzakulov \emph{et al.}
\cite{myrzakulov2011lambdacdm} in the background of $f(G)$ gravity.
Felice and Tsujikawa \cite{de2009construction} explored distinct
constraints for cosmological feasibility of $f(G)$ gravity. Bhatti
\emph{et al.} \cite{bhatti2017evolution,yousaf2017influence,PhysRevD.95.024024,yousaf2019tilted,yousaf2019non} has
explained the circumstances under which anisotropic compact stars
can be formed in modified gravity. Elizalde et al.
\cite{elizalde2020extended} studied the impact of ghost free
$f(R,G)$ model on singular bouncing cosmology.

Here, we intended to study the gravastar under one of these
alternative theories named as $f(R,G)$ gravity and to examine
physical factors of the object. The thumbnail sketch of our paper is
given as. The essential mathematical formalism of the $f(R,G)$
theory has been given in the next section. The field equations in
$f(R,G)$ theory are explained by taking into account the shell,
outer spacetime and inner spacetime of the gravastar in section
\textbf{3}. For the smooth matching of inner and outer regions, we
furnish the required junction conditions in order to make connection
among all sectors of gravastar in section \textbf{4}. In section
\textbf{5}, distinct substantial properties of our model are
described in the scenario of $f(R,G)$ gravity. The whole analysis
has been summarized in the last section.

\section{$f(R,G)$ Gravity}

Here, we will discuss the basic formalism of $f(R,G)$ gravity. We
commence with the action of $f(R,G)$ theory \cite{psaltis2008probes}
which is expressed as
\begin{equation}\label{2}
S=\frac{1}{2k}\int d^{4}x\sqrt{-g}f(R,G)+S_{M}(g^{uv},\psi),
\end{equation}
where $g$ and $S_{M}(g^{\mu\nu},\psi)$ represent the determinant of
metric and matter action respectively and $G$ is the Gauss-Bonnet
invariant. By giving variation to equation ~\eqref{2} with respect
to the metric, it leads to the field equation for $f(R,G)$ theory as
follows
\begin{equation}\label{4}
\kappa
T^{(mat)}_{\mu\nu}+\Sigma_{\mu\nu}=R_{\mu\nu}-\frac{1}{2}g_{\mu\nu}R,
\end{equation}
here $\Sigma_{\mu\nu}$ is given as
\begin{eqnarray}\nonumber
\Sigma_{\mu\nu}&=&2R\nabla_{\nu}\nabla_{\mu}f_{G}+\nabla_{\mu}\nabla_{\nu}f_{R}-g_{\nu\mu}\Box
f_{R}-2g_{\nu\mu}\Box Rf_{G}-4R^{\lambda}_{\mu}
\nabla_{\nu}\nabla_{\lambda}f_{G}\\\nonumber
&-&4R^{\lambda}_{\nu}\nabla_{\mu}\nabla_{\lambda}f_{G}+4R_{\mu\nu}\Box
f_{G}+4g_{\mu\nu}R^{\alpha\beta}\nabla_{\beta}\nabla_{\alpha}f_{G}
+4R_{\mu\alpha\beta\nu}\nabla^{\beta}\nabla^{\alpha}f_{G}\\\label{5}&-&\frac{1}{2}g_{\nu\mu}V+(1-f_{R})(R_{\mu\nu}-\frac{1}{2}g_{\nu\mu}R).
\end{eqnarray}
Here, we have used
\begin{equation}\label{6}
f_{G}\equiv\frac{\partial f}{\partial G},\quad
f_{R}\equiv\frac{\partial f}{\partial R},
\end{equation}
while $V\equiv f_{G}G-f+f_{R}R$ and $T^{(mat)}_{\mu\nu}$ represents
the matter content which for ordinary matter can be written as
\begin{equation}\label{7}
T^{(mat)}_{\nu\mu}=(p+\rho)u_{\nu}u_{\mu}-pg_{\nu\mu},
\end{equation}
here $p$ and $\rho$ indicate the pressure and energy density of the
fluid. Also, the four velocity $u_{\mu}$ satisfies
$u^{\mu}u_{\mu}=1$. The spherically symmetric spacetime inside the
hyper-surface $\Sigma$ is given by
\begin{equation}\label{1}
ds^{2}=e^{\nu}dt^{2}-e^{\lambda}dr^2-r^{2}d\theta^{2}-r^{2}sin^{2}\theta\
d\phi^{2},
\end{equation}
here $\lambda,~\nu$ depend on radial coordinate only.

\section{Field Equations and their Solutions}

In this section, we will evaluate the particular values of the scale
factors for spherical stellar interior which ultimately leads to
gravitational mass of the stellar structure. All of this analysis
would be made by solving modified field equations of the $f(R,G)$
gravity with particular EoS. The non zero components of the Einstein
tensors can be written as
\begin{eqnarray}\label{8}
G_{00}&=&\frac{e^{\nu-\lambda}(-1+e^{\lambda}+\lambda^{'}r)}{r^{2}}
, \\\label{9} G_{11}&=&\frac{1-e^{\lambda}+\nu^{'}r}{r^{2}} ,
\\\label{10}
G_{22}&=&G_{33}=\frac{e^{-\lambda}r\left[2\nu^{'}-2\lambda^{'}+\nu^{'2}r-\nu^{'}\lambda^{'}r+2\nu^{''}r\right]}{4}.
\end{eqnarray}
Here prime is used to show differentiation with reference to radial
coordinate. By using Eqs. (\ref{7})-(\ref{10}) in Eq.(\ref{4}), we
get
\begin{eqnarray}\label{11}
\frac{r^{2}(kT^{0}_{0}+T^{0(D)}_{0})}{e^{-\lambda}f_{R}}&=&e^{\lambda}-1+\lambda^{'}r
, \\\label{12}
\frac{r^{2}(kT^{1}_{1}+T^{1(D)}_{1})}{e^{-\lambda}f_{R}}&=&-1+e^{\lambda}-\nu^{'}r
, \\\label{13}
\frac{r^{2}[kT^{2}_{2}+T^{2(D)}_{2}]}{e^{-\lambda}f_{R}}&=&\left[\frac{r(\lambda^{'}-\nu^{'})}{2}-\frac{r^{2}(2\nu^{"}+\nu^{'2}-\lambda^{'}\nu^{'})}{4}\right].
\end{eqnarray}
Now, with the help of non-conservation equation of effective energy
momentum tensor, we obtain
\begin{eqnarray}\nonumber
&&k\left[e^{-\lambda}\frac{dp}{dr}+\frac{e^{-\lambda}\nu^{'}(\rho+p)}{2}\right]+T^{11(D)'}+\frac{e^{\nu}\nu^{'}T^{00(D)}}{2e^{\lambda}}
+\frac{\lambda^{'}T^{11(D)}}{2}-\frac{rT^{22(D)}}{e^{\lambda}}\\\label{14}&&+\frac{\lambda^{'}T^{11(D)}}{2}+\frac{\nu^{'}T^{11(D)}}{2}
+\frac{2T^{11(D)}}{r}-\frac{rT^{22(D)}}{e^{\lambda}}=0.
\end{eqnarray}
If the gravitational mass of sphere is represented by $m$ then by
using Eq.(\ref{11}), one can have
\begin{align}\label{15}
e^{-\lambda}&=1+\frac{1}{r}\int
r^{2}T^{0(D)}_{0}dr-\frac{2m}{r}f_{R},\quad
\\\label{16}
e^{-\lambda}&=1+\frac{f_{R}}{r}\int
\frac{r^{2}}{f_{R}}T^{0(D)}_{0}dr-\frac{f_{R}}{r}\int
r(1-e^{-\lambda})\left(\frac{1}{f_{R}}\right)'dr-\frac{2m}{r}f_{R}.
\end{align}
It is worthy to mention that Eq.\eqref{15} has been calculated by taking constant values of $R$ and $G$, while Eq.\eqref{16} has radially dependent $R$ and $G$.

\subsection{Interior Spacetime}

Here, we will consider that stellar interior is filled with a
gravitating source. We take EoS for inner sector as taken by Mazur
and Mottola \cite{mazur2004gravitational} as
\begin{equation}\label{17}
p=-\rho.
\end{equation}
The above mentioned EoS is deduced from $p=\omega \rho$ by taking
$\omega=-1$ and is acknowledged as the EoS for dark energy. By
making use of this with Eq.(\ref{14}), we can write
\begin{equation}\label{18}
\rho=-\rho_{0}\quad \textmd{(constant)}
\end{equation}
and pressure becomes
\begin{equation}\label{19}
p=-\rho_{0}.
\end{equation}
For the regular solution at the center we use Eqs.(\ref{11}) and
(\ref{19}) to obtain the value of $\lambda$ with the integration
constant $A=0$ as
\begin{align}\label{21}
e^{-\lambda}&=1-\frac{1}{f_{R}}\left[\frac{K\rho_{0}r^{2}}{3}+r\int
T^{0(D)}_{0}dr-\frac{2}{r}\int (\int
T^{0(D)}_{0}dr)rdr\right],\\\label{23}
e^{-\lambda}&=1-\frac{k\rho_{0}r^{2}}{3f_{R}}+\frac{K\rho_{0}}{3r}\int
r^{3}\left(\frac{1}{f_{R}}\right)^{'}dr-\frac{1}{r}\int
\frac{r^{2}}{f_{R}}T^{0(D)}_{0}dr.
\end{align}
It is worthy to mention that Eq.\eqref{21} has been calculated by taking constant values of $R$ and $G$, while Eq.\eqref{23} has radially dependent $R$ and $G$. We formulate the relation between $\nu$ and $\lambda$ by using Eqs.
(\ref{11}), (\ref{12}), (\ref{18}) and (\ref{19}) as follows
\begin{equation}\label{24}
e^{\nu}=Qe^{-\lambda} ,
\end{equation}
where $Q$ is the integration constant. The mass of gravitating
system $M(D)$ is written as
\begin{align}\label{25}
M(D)&=\frac{1}{f_{R}}\left[\int^{D}_{0}
\frac{r^{2}T^{0(D)}_{0}}{2}dr+\frac{4\pi\rho_{0}D^{3}}{3}\right],\\\label{26} M(D)&=\frac{1}{2}\int^{D}_{0}
\frac{r^{2}T^{0(D)}_{0}}{f_{R}}dr+\frac{4\pi\rho_{0}D^{3}}{3f_{R}}-4\pi\rho_{0}\int^{D}_{0}
\frac{r^{3}}{3}\left(\frac{1}{f_{R}}\right)^{'}dr.
\end{align}
Equation \eqref{25} is found withthe present values of $R$ and $G$, while Eq.\eqref{26} has radially dependent $R$ and $G$.

\subsection{Shell}

We suppose that the shell is made up of ultra-relativistic matter
which implement the EoS as $p=\rho$. A number of researchers studied
different cosmological \cite{madsen1992evolution} and astrophysical
\cite{wesson1978exact,braje2002rx,linares2004importance} events by
using this fluid. It is involuted to simplify the field equations
within the shell. Thus, it is feasible to derive mathematical
solution by taking the limit of thin shell as $1>>e^{-\lambda}>0$.
It implies that the central region should be a thin shell whenever
two spacetimes combine (see \cite{israel1967nuovo}). In shell
$r\rightarrow0$, implies that any parameter having dependence upon
radial coordinate is $<<1 $. With this approximation in addition to
above mentioned EoS coupled with Eqs.(\ref{11})-(\ref{13}), we can
get
\begin{eqnarray}\label{27}
\frac{de^{-\lambda}}{dr}&=&-\frac{r}{f_{R}}\left[T^{0(D)}_{0}+T^{1(D)}_{1}\right]+\frac{2}{r}.
\\\label{28}
\frac{de^{-\lambda}}{dr}\left[\frac{3}{2r}+\frac{\nu^{'}}{4}\right]&=&\frac{1}{r^{2}}-\frac{1}{f_{R}}\left[T^{0(D)}_{0}+T^{2(D)}_{2}\right].
\end{eqnarray}
Integration of Eq.(\ref{27}) yields
\begin{align}\label{30}
e^{-\lambda}&=2\ln r-\frac{1}{f_{R}}\left[r\int
(T^{0(D)}_{0}+T^{1(D)}_{1})dr-\int (\int
(T^{0(D)}_{0}+T^{1(D)}_{1})dr)dr\right]+a_{1}
,\\\label{31} e^{-\lambda}&=2\ln
r-\int \frac{rT^{0(D)}_{0}}{f_{R}}dr-\int
\frac{rT^{1(D)}_{1}}{f_{R}}dr+a_{2},
\end{align}
here $a_{1}$ and $a_{2} $ are integration constants and $r$ having
range $D+\epsilon \geq r \geq D $. The first of the above equations is calculated with the constant values of $R$ and $G$, while the second of the above equations has radially dependent $R$ and $G$. Due to the condition $\epsilon <<
1$ and $e^{-\lambda}<< 1$, we obtain $a_{1}<<1$ and $a_{2}<<1 $.
From Eqs.(\ref{27}) and (\ref{28}), one can write
\begin{equation}\label{32}
e^{\nu}=Ze^{B} ,
\end{equation}
where $Z$ is the constant of integration. Equation (\ref{14}) as
well as the EoS $p=\rho$ results into
\begin{equation}\label{33}
p=\rho=He^{-B+\int T^{(D^{\star})}dr}.
\end{equation}
\begin{figure} \centering
\epsfig{file=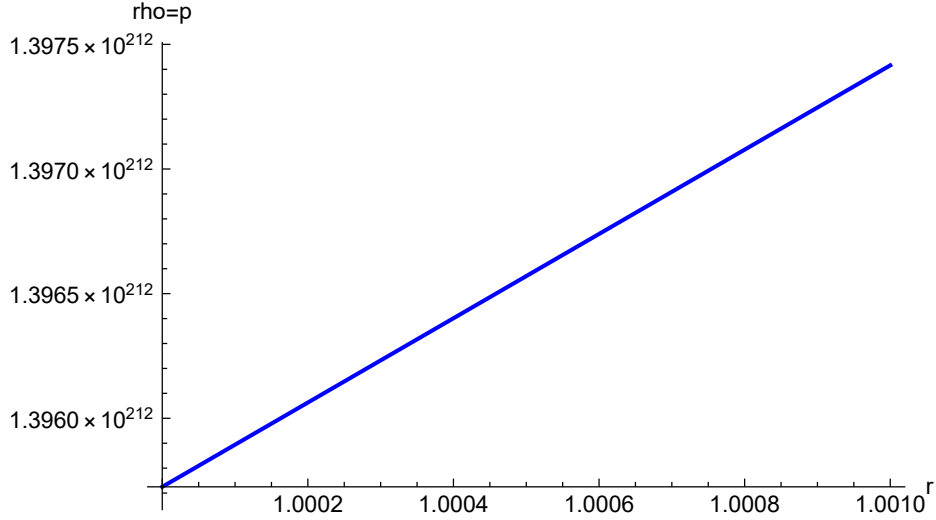,width=0.9\linewidth}
\caption{graphical representation of density relationship with r}
\end{figure}

\subsection{Exterior Spacetime}

The outer sector is described with the help of widely known static
exterior Schwarzschild solution whose mathematical form is written
as
\begin{equation}\label{34}
ds^{2}=-\left(1-\frac{2M}{r}\right)^{-1}dr^{2}-r^{2}(d\theta^{2}-\sin^{2}\theta
d\phi^{2})+\left(1-\frac{2M}{r}\right)dt^{2}.
\end{equation}
where $M$ represents the mass of gravitational system.

\section{ Junction Condition}

The interior sector (I) of gravastar is coupled with the exterior
sector (III) at the shell. Gravastar must have flat matching in the
middle of sectors (I) and (III) as mentioned by the Israel formalism
\cite{darmois1927memorial,israel1966singular}. At the coupling
surface, the metric is continuous nevertheless they may not have
continuous derivatives. We can find out $S_{ij}$ by implementing the
above stated formalism. Lanczos equation
\cite{lanczos1924flachenhafte,sen1924grenzbedingungen,perry1992traversible,musgrave1996junctions}
states $S^{i}_{j}$ as
\begin{equation}\label{35}
S^{i}_{j}=\frac{1}{8\pi}\left(\delta^{i}_{j}\kappa^{k}_{k}-\kappa^{i}_{j}\right),
\end{equation}
here $k_{ij}=K^{+}_{ij}-K^{-}_{ij}$ yield the discontinuity in the
extrinsic curvatures. The inner and outer sectors are correlated by
``+'' and ``-'' signs, respectively. The second fundamental forms
\cite{rahaman2006thin,usmani2010thin,dias2010thin,rahaman2011comparison}
linked with the two borders of the shell are given by
\begin{equation}\label{36}
K^{\pm}_{ij}=-n^{\pm}_{\nu}\left[\Gamma^{\nu}_{\alpha\beta}\frac{\partial
x^{\alpha}}{\partial \xi^{i}}\frac{\partial x^{\beta}}{\partial
\xi^{j}}+\frac{\partial^{2}x_{v}}{\partial\xi^{i}\partial\xi^{j}}\right]_{\Sigma}
,
\end{equation}
where $\xi^{i}$ and $n^{\pm}_{\nu}$ represent the intrinsic
coordinates of the shell and unit normal over $\Sigma$,
respectively. The line element for spherical geometry is
\begin{equation}\label{37}
ds^{2}=-\frac{dr^{2}}{f(r)}+f\left(r\right)d^{2}-r^{2}(d\theta^{2}-\sin^{2}\theta)
d\phi^{2} ,
\end{equation}
for which the unit normal $n^{\pm}_{\nu}$ can be written as
\begin{equation}\label{38}
n^{\pm}_{\nu}=\pm\mid g^{\alpha\beta}\frac{\partial f}{\partial
x^{\alpha}}\frac{\partial f}{\partial x^{\beta}}\mid^{\frac{-1}{2}}\frac{\partial f}{\partial x^{\nu}} ,
\end{equation}
with $n^{\mu}n_{\mu}$=1. With the help of Lanczos equation, one can
write  $S_{ij}=diag[\sigma,-\mathbf{\nu}]$, where $\nu$ is written
for the surface pressure and $\sigma$ represents surface energy
density. The density distribution as well as pressure at the surface
are written as
\begin{eqnarray}\label{39}
\sigma&=&-\frac{1}{4\pi D}\left[\sqrt{f}\right]^{+}_{-},
\\\label{40}
\nu&=&-\frac{\sigma}{2}+\frac{1}{16\pi}\left[\frac{f^{'}}{\sqrt{f}}\right]^{+}_{-}.
\end{eqnarray}
By making use of the above two equations when $e^{-\lambda}=L$ and
$e^{-\lambda}=A$,  respectively, we get
\begin{eqnarray}\label{41}
\sigma&=&\frac{-1}{4\pi
D}\left[\sqrt{1-\frac{2M}{D}}-\sqrt{L}\right],\\\label{42}
\sigma&=&\frac{-1}{4\pi
D}\left[\sqrt{1-\frac{2M}{D}}-\sqrt{A}\right],\\\label{43}
\upsilon&=&\frac{1}{8\pi
D}\left[\frac{1-\frac{M}{D}}{\sqrt{1-\frac{2M}{D}}}-\frac{2f_{R}D(L)-M}{2f_{R}D\sqrt{L}}\right],\\\label{44}
\upsilon&=&\frac{1}{8\pi
D}\left[\frac{1-\frac{M}{D}}{\sqrt{1-\frac{2M}{D}}}-\frac{A+B-C+E}{\sqrt{A}}\right].
\end{eqnarray}
By using Eq.(\ref{41}), the thin shell mass is obtained as
\begin{eqnarray}\nonumber
m_{s}&=&D\left[\sqrt{1-\frac{1}{f_{R}}\left[\frac{k\rho_{0}D^{2}}{3}
+D\int T^{0(D)}_{0}dD-\frac{2}{D}\int (\int T^{0(D)}_{0}dD)DdD\right]}\right.\\\label{45}&-&\left.\sqrt{1-\frac{2M}{D}}\right].
\end{eqnarray}
In a similar ways, from Eq.(\ref{42}), the thin shell mass is
obtained as
\begin{equation}\label{46}
m_{s}=D\left[\sqrt{1-\frac{k\rho_{0}D^{2}}{3f_{R}}+\frac{k\rho_{0}}{3D}
\int D^{3}(\frac{1}{f_{R}})^{'}dD-\frac{1}{D}\int \frac{D^{2}}{f_{R}}T^{0(D)}_{0}dD}-\sqrt{1-\frac{2M}{D}}\right].
\end{equation}
By using Eq.(\ref{43}), the total mass $M$ of gravastar is found to
be
\begin{equation}\label{47}
M=\frac{D}{2}-\frac{m_{s}^{2}}{2D}-\frac{D}{2}\left[L\right]+m_{s}\sqrt{L},
\end{equation}
while from Eq.(\ref{44}), the total mass $M$ of gravastar is written
as
\begin{equation}\label{48}
M=\frac{-m^{2}_{s}}{2D}+\frac{D}{2}\left[1-A\right]+m_{s}\sqrt{A}.
\end{equation}

\section{Realistic Characteristics}

This section is aimed to examine the impact of $f(R,G)$ gravity on
various physical characteristics of gravastar model. In particular,
we will analyze the shell's length as well as energy of the stellar
model. The entropy as well as EoS will also be discussed during the
dynamical formulation of gravastars. These results would also be
indicated via plots.

\subsection{Proper length of the shell}

We assume that shell is placed at $r=D$ which defines the phase edge
of sector (I). The shell is of very small width, i.e., $\epsilon
<<1$. Therefore, the sector (III) initiates from attachment at
$r=D+\epsilon$. Thus, the actual width in between of these
attachments of the shell is calculated by using Eq.(\ref{30}) as
\begin{equation}\label{49}
\ell=\int^{D+\epsilon}_{D}\frac{dr}{\sqrt{2lnr+c_{1}-\frac{1}{f_{R}}
\left[r\int (T^{0(D)}_{0}+T^{1(D)}_{1})dr-\int (\int (T^{0(D)}_{0}+T^{1(D)}_{1})dr)dr\right]}}.
\end{equation}
\begin{figure} \centering
\epsfig{file=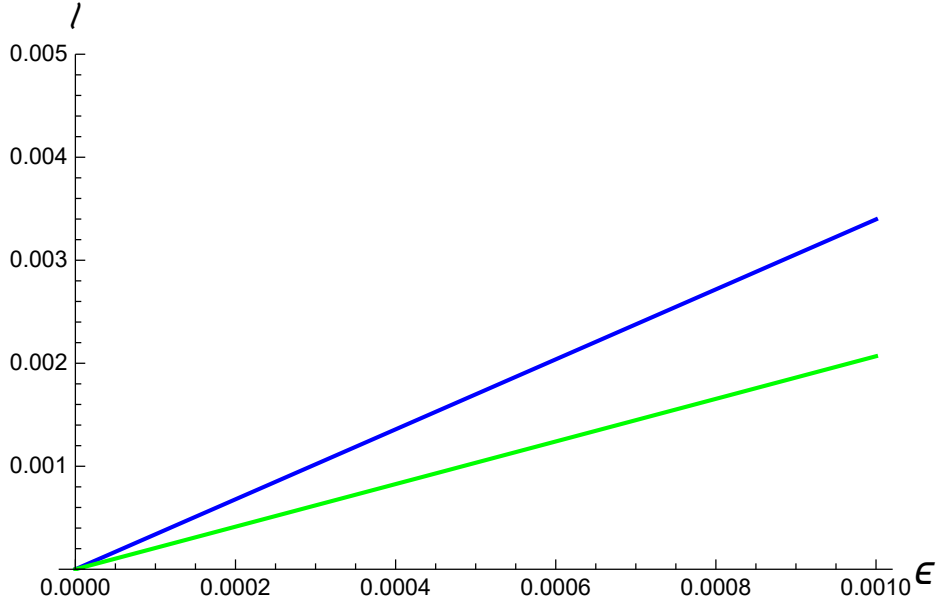,width=0.9\linewidth}
\caption{graphical representation of relationship between proper length $\ell$ and the thickness of shell $\epsilon$}
\end{figure}
By making use of Eq.(\ref{31}), one can
have\begin{equation}\label{50}
\ell=\int^{D+\epsilon}_{D}\frac{dr}{\sqrt{2lnr+c_{2}-\int
\left(\frac{r}{f_{R}}T^{0(D)}_{0}\right)dr-\int
\frac{r}{f_{R}}T^{1(D)}_{1}dr}}.
\end{equation}
\begin{figure} \centering
\epsfig{file=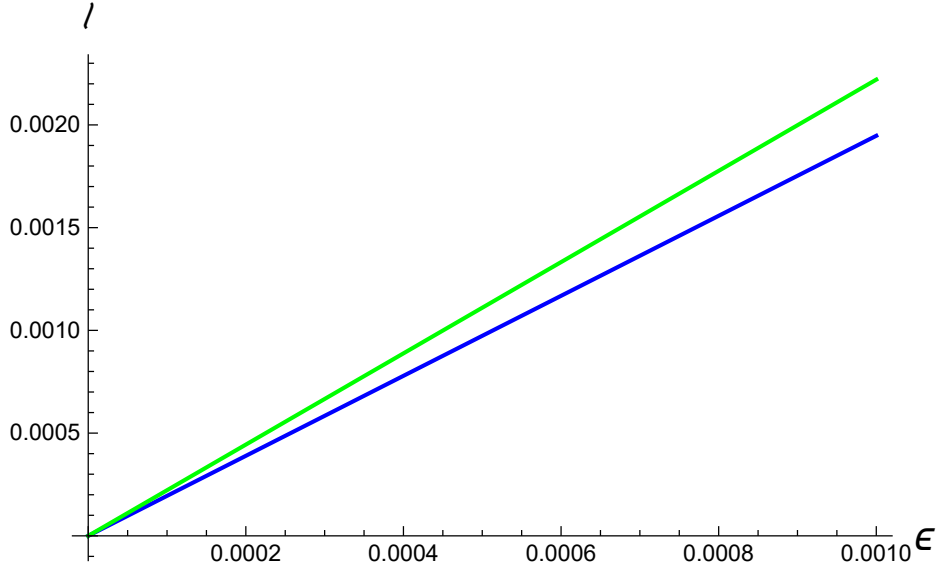,width=0.9\linewidth}
\caption{graphical representation of relationship between proper length $\ell$  and the thickness of shell $\epsilon$}
\end{figure}

\subsubsection{Energy content}

We suppose the EoS $(p=-\rho)$ for the inner sector that illustrates
the zone having negative energy and approve the nature of repulsion
of interior sector. But the energy inside the shell will be
\begin{equation}\label{51}
\varepsilon=\int^{D+\epsilon}_{D} 4\pi\rho r^{2}dr=4\pi
H\int^{D+\epsilon}_{D} e^{-B+\int T(D^{\star})dr}r^{2}dr.
\end{equation}
\begin{figure} \centering
\epsfig{file=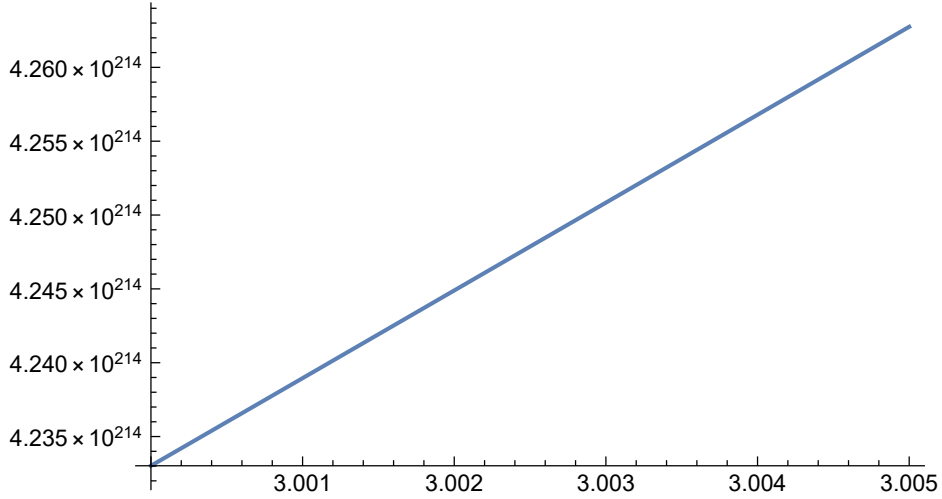,width=0.9\linewidth}
\caption{graphical representation of energy $\varepsilon$ relationship with thickness of shell $\epsilon$}
\end{figure}

\subsubsection{Entropy}

Mazur and Mottola \cite{mazur2004gravitational} suggested that the
interior sector must has zero entropy that is persistent with a
specific condensate condition. Inside the shell, the entropy will be
\begin{equation}\label{52}
S=\int^{D+\epsilon}_{D}4\pi r^{2}s(r)\sqrt{e^{\lambda}}dr.
\end{equation}
The entropy of local temperature $T(r)$ is given by
\begin{equation}\label{53}
S(r)=\frac{\alpha^{2}k^{2}_{B}T(r)}{4\pi
h^{2}}=\alpha\left(\frac{k_{B}}{h}\right)\sqrt{\frac{p}{2\pi}} ,
\end{equation}
where $\alpha$ is a constant having no dimension and $G=c=1$ with
Planckian units $k_{B}=h=1$. The entropy density inside the shell
will be
\begin{equation}\label{54}
S(r)=\alpha\sqrt{\frac{p}{2\pi}}.
\end{equation}
So, by using Eq.(\ref{30}) in (\ref{52}), we get
\begin{equation}\label{55}
S=\sqrt{8\pi H}\alpha\int^{D+\epsilon}_{D}r^{2}\sqrt{e^{-B+\int T(D^{\star})dr}}\frac{dr}{\sqrt{Z}}.
\end{equation}
In a similar manner, the use of Eq.(\ref{31}) in (\ref{52}) leads to
\begin{equation}\label{56}
S=\sqrt{8\pi H}\alpha\int^{D+\epsilon}_{D}r^{2}\sqrt{e^{-B+\int T(D^{\star})dr}}\frac{dr}{\sqrt{T
}}.
\end{equation}

\subsubsection{Equation of State}

In general, the EoS at $r=D$ is given as
\begin{equation}\label{57}
\upsilon=\omega(D)\sigma
\end{equation}
By making use of Eqs.(\ref{41}) and (\ref{43}), we found
\begin{equation}\label{58}
\omega(D)=\frac{\left[\frac{1-\frac{M}{D}}{\sqrt{1-\frac{2M}{D}}}-\frac{2f_{R}DL-M}{2f_{R}D\sqrt{L}}\right]}{2\left[-\sqrt{1-\frac{2M}{D}}+\sqrt{L}\right]},
\end{equation}
here, $\omega(D)$ can only be real if $\frac{2M}{D}<1$ along with
$$\frac{1}{f_{R}}\left[\frac{k\rho_{0}D^{2}}{3}+D\int T^{0(D)}_{0}dD
-\frac{2}{D}\int (\int T^{0(D)}_{0}dD)DdD\right]<1.$$ Also, by
taking $\frac{M}{D}<<1$ and
$\frac{1}{f_{R}}\left[\frac{k\rho_{0}D^{2}}{3}+D\int
T^{0(D)}_{0}dD-\frac{2}{D}\int (\int T^{0(D)}_{0}dD)DdD\right]<<1$
in a binomial series and having the terms up to first order for the
terms in the square-root, we obtain
\begin{equation}\label{59}
\omega(D)\approx\frac{1-L+\frac{D}{2f_{R}}\int T^{0(D)}_{0}dD-\frac{1}{f_{R}D}\int\int ( T^{0(D)}_{0}dD)DdD}{2\left[\frac{M}{D}-\frac{1}{2}(1-L)\right]}.
\end{equation}
By making use of Eqs.(\ref{42}) and (\ref{44}), one can write
\begin{equation}\label{60}
\omega(D)=\frac{\frac{1-\frac{M}{D}}{\sqrt{1-\frac{2M}{D}}}-\frac{A+B-C-D}{\sqrt{A}}}{2\left[\sqrt{1-\frac{2M}{D}+\sqrt{A}}\right]}.
\end{equation}
Here, $\omega(D)$ can only be real if \quad$\frac{2M}{D}<1$ along
with \quad$\frac{K\rho_{0}D^{2}}{3f_{R}}+\frac{K\rho_{0}}{3D}\int
D^{3}(\frac{1}{f_{R}})^{'}dD+\frac{1}{D}\int
\frac{D^{2}}{f_{R}}T^{0(D)}_{0}dD<1$. Also, if one elaborate the
square-root terms of Eq.(\ref{61}) by taking $\frac{M}{D}<<1$ and
$\frac{K\rho_{0}D^{2}}{3f_{R}}+\frac{K\rho_{0}}{3D}\int
D^{3}(\frac{1}{f_{R}})^{'}dD+\frac{1}{D}\int
\frac{D^{2}}{f_{R}}T^{0(D)}_{0}dD<<1$ in a binomial series and
having the terms up to first order, then we get
\begin{equation}\label{61}
\omega(D)\approx{\frac{\frac{2K\rho_{0}D^{2}}{3f_{R}}
+\frac{D^{2}}{2f_{R}}T^{0(D)}_{0}+....}{2\left[\frac{M}{D}-\frac{K\rho_{0}D^{2}}{6f_{R}}+....\right]}}.
\end{equation}

\section{Conclusion}

In this paper, we have demonstrated a particular stellar model which
was initially theorized by Mazur and Mottola
\cite{mazur2004gravitational} under the influence of $f(R,G)$
gravity. The stellar model named as gravastar can be studied as a
feasible alternate to the black hole structure. The gravastar can be
characterized by three distinct sectors; inner sector, intermediate
thin shell and outer sector with particular EoS for every sector.
With the help of this description, we have calculated a definite
solution which is free of singularity for the gravastar and
demonstrated it with different physically feasible properties within
the scheme of $f(R,G)$ gravity.

We have written down distinct important features of the solution set
as follows:
\begin{enumerate}
  \item Density-pressure relationship: The relation of pressure of the ultra-relativistic
  matter present in the shell with its density is shown corresponding to its radius in Fig.\textbf{1}
  which sustains a consistent diversity in every part of the shell.
  \item Proper length: Relationship between proper length $\ell$ and thickness of shell ($\epsilon$)(in Fig.\textbf{2})
  shows the continuous increase (constant case).
  \item Proper length: Relationship between proper length $\ell$ and thickness of shell ($\epsilon$)
  (in Fig.\textbf{3}) shows the continuous increase (variable case).
  More precisely, the
  influence of $f(R,G)$ is responsible to increase the length of the
  shell.
  \item Energy content: Figure \textbf{4} shows that the energy of the shell has direct relation corresponding to the thickness of the shell $\epsilon$.
\end{enumerate}

\noindent\textbf{Declaration of Competing Interest}\\

The authors declare that they have no known competing financial
interests or personal relationships that could have appeared to
influence the work reported in this paper.\\

\noindent\textbf{Acknowledgement}\\

The work of MZB and ZY was supported by National Research Project
for Universities (NRPU), Higher Education Commission, Pakistan under
research project No. 8769/Punjab/ NRPU/R$\&$D/HEC/2017. The authors
would like to thank the anonymous reviewer for the valuable and
constructive comments and suggestions in order to improve the
quality of the paper.\\

\section{Appendix:}

\begin{align}\nonumber
L&=1-\frac{1}{f_{R}}\left[\frac{k\rho_{0}D^{2}}{3}+D\int T^{0(D)}_{0}dD-\frac{2}{D}\int (\int T^{0(D)}_{0}dD)DdD\right],
\\\nonumber
M&=2K\rho_{0}D^{3}+D^{2}\int T^{0(D)}_{0}dD+D^{3}T^{0(D)}_{0}-2D^{2}\int T^{0(D)}_{0}dD\\\nonumber
&+2\int (\int T^{0(D)}_{0}dD)DdD,
\\\nonumber
A&=1-\frac{K\rho_{0}D^{2}}{3f_{R}}+\frac{K\rho_{0}}{3D}\int D^{3}\left(\frac{1}{f_{R}}\right)^{'}dD+\frac{1}{D}\int \frac{D^{2}}{f_{R}}T^{0(D)}_{0}dD,
\\\nonumber
B&=\frac{K\rho_{0}D}{6}\left[\frac{-2Df_{R}+D^{2}(f_{R})^{'}}{(f_{R})^{2}}\right],
\\\nonumber
C&=\frac{K\rho_{0}D}{6}\left[\frac{-1}{D^{2}}\int D^{3}\left(\frac{1}{f_{R}}\right)^{'}dD+D^{2}\left(\frac{1}{f_{R}}\right)^{'}\right],
\\\nonumber
T&=2lnr-\int \frac{r}{f_{R}}(T^{0(D)}_{0}+T^{1(D)}_{1})dr+c_{6},
\\\nonumber
E&=\frac{D}{2}\left[\frac{-1}{D^{2}}\int \frac{D^{2}}{f_{R}}T^{0(D)}_{0}dD+\frac{D}{f_{R}}T^{0(D)}_{0}\right],
\\\nonumber
Z&={2lnr-\frac{1}{f_{R}}[r\int(T^{0(D)}_{0}+T^{1(D)}_{1})dr-\int (\int (T^{0(D)}_{0}+T^{1(D)}_{1}dr)dr)]+c_{5}},
\\\nonumber
T^{(D^\star)}&=\frac{1}{PKe^{-\lambda}}\left[-T^{11(D)'}-\frac{e^{\nu}\nu'}{2e^{\lambda}}T^{00(D)}
-\lambda'T^{11(D)}+\frac{r}{e^{\lambda}}T^{22(D)}-\frac{\nu'}{2}T^{11(D)}\right.\\\nonumber
&-\left.\frac{2}{r}T^{11(D)}+\frac{r}{e^{\lambda}}T^{22(D)}\right].
\end{align}

\end{document}